\documentclass[]{amsart}
\usepackage{amsmath, amssymb, stmaryrd, amsthm}
\usepackage[all,knot]{xy}
\usepackage{tikz}
\usepackage{url,verbatim,hyperref}

\newcommand{\B}{\mathcal{B}}
\newcommand{\LB}{\mathcal{LB}}

\newcommand{\R}{\mathbb{R}}

\newcommand{\mC}{\mathcal{C}}

\newcommand{\mZ}{\mathcal{Z}}

\newcommand{\mcc}{\mathcal{C}}

\newcommand{\bcg}{\underline{Aut_{\otimes}^{br}(\mathcal{C})}}
\newcommand{\Pic}{\underline{\textrm{Pic}(\mcc)}}

\numberwithin{equation}{section}

\theoremstyle{definition}
\newtheorem{definition}{Definition}[section]

\theoremstyle{definition}

\theoremstyle{remark}

\begin{document}

\title{Beyond Anyons}
\date{\today}

\author{Zhenghan Wang}
\email{zhenghwa@microsoft.com}
\address{Microsoft Research Station Q and Department of Mathematics,
    University of California,
    Santa Barbara, CA
    U.S.A.}

\begin{abstract}

The theory of anyon systems, as modular functors topologically and unitary modular tensor categories algebraically, is mature.  To go beyond anyons, our first step is the interplay of anyons with conventional group symmetry due to the paramount importance of group symmetry in physics.  This led to the theory of symmetry-enriched topological order.  Another direction is the boundary physics of topological phases, both gapless as in the fractional quantum Hall physics and gapped as in toric code.
A more speculative and interesting direction is the study of Banados-Teitelboim-Zanelli black holes and quantum gravity in $3d$.  The clearly defined physical and mathematical issues require a far-reaching generalization of anyons and seem to be within reach.
In this short survey, I will first cover the extensions of anyon theory to symmetry defects and gapped boundaries.  Then I will discuss a desired generalization of anyons to anyon-like objects---the Banados-Teitelboim-Zanelli black holes---in $3d$ quantum gravity.

\end{abstract}
\thanks{Partially supported by NSF DMS grant 1411212}
\maketitle

\section{Introduction}

Systematic application of quantum topology in condensed matter physics accelerated significantly after the $2003$ workshop {\it topological phases in condensed matter physics} \cite{FNW03}.  Anyons, elementary excitations in $2D$ topological phases of matter, play a central role in this new period of interactions between topology and physics in $3d$ space-time\footnote{I will use the convention that $nD$ means the $D$-dimensional space and $nd$ the $d$-dimensional space-time, so $(n+1)d=nD+1$.} (see \cite{Wangbook, RWsurvey} and the references therein).  Conceptually, anyons provide an interpolation between bosons and fermions through the topological spins of abelian anyons: $e^{i\theta}$ for some $\theta$'s with $\theta=0$ bosons and $\theta=\pi$ fermions.  I see a striking parallel of the application of topology in physics to the application of topology in differential geometry---global differential geometry.  Progress in science and technology has made topological physics inevitable: the inherently discrete nature of quantum mechanics, the continuing miniaturization of quantum devices, and the maturity of local classical physics.

At the forefront of topological physics is the application to quantum computing: topological quantum computing with the promise of an inherently fault-tolerant universal quantum computer \cite{FKLW}.  However, to build a real topological quantum computer, topological physics has to be coupled to conventional physics such as during the initialization and read-out.  Therefore, it is natural to consider how to extend topological physics beyond anyons.  One obvious direction is from $2D$ to $3D$.  But the difference between $2D$ and $3D$ could be enormous as the topology of $3d$ and $4d$ manifolds is dramatically different.  For example, we have a rather complete classification of $3d$ space-time manifolds due to the geometrization theorem of Pereleman-Thurston, while there is no reasonable conjectured picture of smooth $4d$ space-time manifolds.

The theory of anyon systems, as modular functors topologically and unitary modular tensor categories algebraically, is mature \cite{RWsurvey}.  To go beyond anyons, our first step is the interplay of anyons with conventional group symmetry due to the paramount importance of group symmetry in physics.  This led to the theory of symmetry-enriched topological order.  Another direction is the boundary physics of topological phases, both gapless as in the fractional quantum Hall physics and gapped as in toric code.

A more speculative and interesting direction is the study of Banados-Teitelboim-Zanelli (BTZ) black holes and quantum gravity in $3d$.  The clearly defined physical and mathematical issues require a far-reaching generalization of anyons and seem to be within reach.

In this survey, I will first cover the extensions of anyon theory to symmetry defects and gapped boundaries.  Then I will discuss a desired generalization of anyons to anyon-like objects---the BTZ black holes---in $3d$ quantum gravity.

\section{$2D$ Non-abelian Objects}

Non-abelian means the order in a sequence of things is important such as the order of letters in words: NO is not the same as ON.  If many non-abelian objects $X_1,..X_n$ are lined up in a line, then their states can be changed by exchanging any two of them.  If two different exchanges are performed sequentially, the order of the two exchanges becomes important.  The fundamental prerequisite for such a phenomenon is that the states of such $n$ non-abelian objects are not unique, i.e. the ground states have degeneracy, which is more fundamental than statistics \cite{RWdegeneracy}.  The best understood non-abelian objects are non-abelian anyons.  The last two decades of research in condensed matter physics has yielded remarkable progress in the understanding of non-abelian anyons in topological phases of matter.  Recently, other non-abelian objects are discovered such as the Majorana zero modes, which lead essentially to the same non-abelian physics as non-abelian anyons.  These new non-abelian objects---symmetry defects and gapped boundaries---open doors to new approaches to topological quantum computation.

\subsection{Symmetry defects}

In the absence of any symmetry, gapped quantum systems at zero temperature can still form distinct phases of
matter---topological phases of matter (TPMs), which are characterized by their topological order.

A TPM $\mathcal{H}=\{H\}$ is an equivalence class of gapped Hamiltonians $H$ which realizes a TQFT at low energy.  Elementary excitations in a TPM $\mathcal{H}$ are point-like anyons.  Anyons can be modeled algebraically as simple objects in a unitary modular tensor category (UMC) $\mathcal{B}$, which will be referred to as the topological order of the TPM $\mathcal{H}$.

The interplay of symmetry with topological order has generated intense research. In the presence of symmetries, TPMs acquire a
finer classification and fractional quasi-particles
of a topologically ordered state can acquire fractional quantum numbers of the global symmetry.

When a Hamiltonian for a TPM possesses a global symmetry, it is natural
to consider the topological order that is obtained when this global symmetry is promoted to a
local, gauge symmetry. This gauging procedure is useful in many ways.

\subsection{Algebraic model of symmetry defects}

In the real world, TPMs are always coupled to conventional degrees of freedom.  TPMs with conventional group symmetries are called symmetry enriched topological phases of matter (SETs). When the intrinsic topological order is trivial, SETs become symmetry protected topological phases (SPTs).  Important examples of SPTs are topological insulators and topological superconductors.

Let $G$ be a finite group and $\mcc$ a UMC, also called an anyon model.

\subsubsection{Topological Symmetry}
Promoting $G$ to a categorical-group $\underline{G}$, we denote by $\underline{Aut_{\otimes}^{br}(\mathcal{C})}$ the categorical-group of braided tensor autoequivalences of the UMC $\mathcal{C}$, which is the full topological symmetry of $\mcc$.

\begin{definition}

A finite group $G$ is a topological symmetry of the UMC $\mcc$ if there is a monoidal functor $\underline{\rho}: \underline{G}\rightarrow \bcg$.
The topological symmetry is denoted as $(\underline{\rho}, G)$ or simply $\underline{\rho}$.

\end{definition}

\subsubsection{Symmetry Defects}

The invertible module categories over $\mcc$ form the Picard categorical-group $\underline{\textrm{Pic}(\mcc)}$ of $\mcc$.  The Picard categorical-group $\underline{\textrm{Pic}(\mcc)}$ of the UMC $\mcc$ is monoidally equivalent to the categorical-group $\bcg$.  This one-one correspondence between braided tensor auto-equivalences and invertible module categories underlies the relation between symmetry and defect.  Hence, given a topological symmetry $(\underline{\rho}, G)$ of the UMC $\mcc$ and an isomorphism of $\Pic$ with $\bcg$, each $\underline{\rho}(g)\in \bcg$ corresponds to an invertible bi-module category $\mcc_g\in \Pic$.

\begin{definition}

An extrinsic topological defect of flux $g\in G$ is a simple object in the invertible module category $\mathcal{C}_g\in \underline{\textrm{Pic}(\mcc)}$ corresponding to the braided tensor autoequivalence $\underline{\rho}(g)\in \bcg$.

\end{definition}

\subsubsection{Gauging Topological Symmetry}

Let $\underline{\underline{G}}$ be the categorical $2$-group, and $\underline{\bcg}$ be the categorical $2$-group of braided tensor auto-equivalences.

\begin{definition}

A topological symmetry $\underline{\rho}: \underline{G}\rightarrow \bcg$ can be gauged if $\underline{\rho}$ can be lifted to a categorical $2$-group functor $\underline{\underline{\rho}}: \underline{\underline{G}} \rightarrow \underline{\bcg}$.

\end{definition}

The physical and mathematical theory of symmetry defects can be found in \cite{BBCW,CCPW}.
Symmetry defects can be used to enhance the computing power of anyons.  Study of topological quantum computation with symmetry defects can be found in \cite{DW17}.

\subsection{Gapped boundaries}

A second direction for non-abelian objects beyond anyons is gapped boundaries in TPMs which are Drinfeld centers $\mathcal{Z}(\mC)$ of unitary fusion categories $\mC$, called doubled theory in physics.

The Levin-Wen model (LW) in $2D$ is a lattice Hamilotnian realization of Turaev-Viro type TQFTs based on unitary fusion categories.  The conceptual underpinning of the $2D$ LW model is two mathematical theorems:  The Drinfeld center $\mZ(\mC)$ of a unitary fusion category $\mC$ is always modular, and the Turaev-Viro TQFT based on $\mC$ is equivalent to the Reshetikhin-Turaev TQFT based on $\mZ(\mC)$.  Therefore, LW model is a lattice Hamiltonian implementation of both theorems simultaneously.  Those rigorously solvable models provide the best playground for the theoretical study of TPMs.  Realistically, samples of TPMs have boundaries.  The interplay of the boundary with the interior or bulk contains rich physics as exemplified by the famous holographic principle.

In the categorical formalism, the bulk of a doubled TQFT is given by a UMC $\mathcal{B} = \mathcal{Z}(\mC)$ for some unitary fusion category $\mC$, and a (gapped) hole is a Lagrangian algebra $\mathcal{A}=\oplus_{a}n_a a$ in $\mathcal{B}$. In the case of Dijkgraaf-Witten theories, we have $\mC = \text{Vec}_G$. For most purposes, $\mathcal{A}$ can be regarded as a (composite) non-abelian anyon of quantum dimension $d_{\mathcal{A}}$. Gapped boundaries are conjectured to be in one-to-one correspondence to indecomposable module categories $\mathcal{M}_i$ over $\mC$. Then, elementary excitations on $\mathcal{M}_i$ are the simple objects in the functor fusion category $\mC_{ii} = \textrm{Fun}_{\mC}(\mathcal{M}_i, \mathcal{M}_i)$, and simple
boundary defects between two gapped boundaries $\mathcal{M}_i, \mathcal{M}_j$ are the simple objects in the bimodule category  $\mC_{ij} = \textrm{Fun}_{\mC}(\mathcal{M}_i, \mathcal{M}_j)$.  The collections of fusion categories $\mC_{ii}$ and their bimodule categories $\mC_{ij}$ form a multi-fusion category $\mathfrak{C}$. From this multi-fusion category, we can find quantum dimensions of both boundary excitations and the defects between gapped boundaries.

Topological quantum computation with gapped boundaries is investigated in \cite{CCW16} and a striking example is the universal gate set from a purely abelian TPM \cite{CCW17}.  Topological quantum computation with boundary defects are also very interesting, but a systematical study has not been initiated.

\section{Three dimensional topological physics beyond anyons}

Classically the universe has three spacial dimensions, but nano-technology makes the study of low dimensional physics exciting such as anyons in $2D$.  Therefore both as a toy model for $3D$ physics and potentially realistic low dimensional physics, it is interesting to consider all possible $3d$ physics including the Yang-Mills theory.  One salient feature of $3d$ is the Chern-Simons (CS) action which could be coupled to Yang-Mills.  A fascinating direction is $3d$ quantum gravity.  Classical $3d$ gravity is the same as a doubled CS theory with gauge group $SL(2,\R)$, but how to quantize doubled $SL(2,\R)$-CS theory, which has a non-compact gauge group, is very challenging.  This is an excellent example for the generalization of anyon systems to topological systems with infinitely many elementary excitation types, closely related to Banados-Teitelboim-Zanelli (BTZ) black holes.  The geometry, topology, and physics in and around $3d$ pure quantum gravity with negative cosmological constant center on the relation between quantum $3D$ gravity and quantum doubled CS gauge theory (complicated by the invertibility of viebeins), the existence of BTZ black holes, and the asymptotic Virasoro algebra discovered by Brown and Henneaux, which is a precursor of the Ads/CFT correspondence.

\subsection{3d gravity}

Let $X^3$ be a closed oriented $3d$ space-time manifold and $g$ a gravitational field.  The Einstein-Hilbert action is
$$I(g)=\int_{X^3} d^3x \sqrt{g} (R-2\Lambda),$$
where $R$ is the scalar curvature and $\Lambda$ the cosmological constant.  The equation of motion gives rise to the Einstein equation:
$$R_{\mu \nu}-\frac{1}{2}R g_{\mu\nu}+\Lambda g_{\mu\nu}=0,$$
where $R_{\mu\nu}$ is the Ricci curvature.

The $3d$ anti-de Sitter space is the subspace of $\R^4$ defined as
$$\{ -x_1^2-x_2^2+x_3^2+x_4^2=-l^2 \}$$
for some constant $l>0$ with the metric $ds^2= -dx_1^2-dx_2^2+dx_3^2+dx_4^2$.
Direct computation gives
$$R_{\mu \nu}=-\frac{2}{l^2}g_{\mu\nu}, R=-\frac{6}{l^2},$$ therefore the gravitational field $ds^2= -dx_1^2-dx_2^2+dx_3^2+dx_4^2$
is a solution of the Einstein equation if we choose the negative cosmological constant $\Lambda=-\frac{1}{l^2}$.
Topologically the anti-de Sitter space is simply $S^1\times \R^2$.

The tangent bundle of $X^3$ is trivial so $TX\cong X^3\times \R^3$.  A framing $e: TX\cong X^3\times \R^3$ is a choice of such an identification, which is called a vierbein (really dreibein) in physics.  Let $\omega$ be a spin conenction, then $(\omega, e)$ can be made into an $SO(2,2)$ gauge field.  Let $A_{\pm}=\omega \pm \frac{e}{l}$, then the Einstein-Hilbert action becomes a doubled CS action $I(X,g)=\frac{k_L}{4\pi}CS(A_+)-\frac{k_R}{4\pi}CS(A_{-})$.  Therefore, classical $3d$ gravity with $\Lambda <0$ is the same as doubled CS theory with levels $k_L=k_R=\frac{3G}{2l}$, where $G$ is the Newton constant \cite{Witten88}.

The quantum theory of $3d$ gravity is more subtle as the correspondence with CS theory is not exact due to the difference between gauge transformations and non-invertible vierbeins.
But if $3d$ quantum gravity can be defined mathematically, it should be some irrational TQFT.  Then solving $3d$ quantum gravity would be to find the corresponding conformal field theory in a sense, presumably irrational too.  Speculatively, then BTZ back holes would be anyon-like objects.

\subsection{Volume conjecture}

One profound implication of $3d$ quantum gravity is the possibility of a volume conjecture \cite{Witten88}.  The volume conjecture is made precise for hyperbolic knots \cite{KVolumeknot}.  Our interest is on closed hyperbolic $3$-manifolds.  It can be easily seen that the naive generalization from knots to closed $3$-manifolds using rational unitary TQFTs cannot be correct as the $3$-manifold invariants grow only polynomially as the level goes to infinity.  The most promising version is to use non-unitary rational TQFTs as in \cite{CY}.
It is puzzling how the non-unitarity arises from the unitary $3d$ quantum gravity.

\section{Four dimensional topological physics}

Two interesting families of $(3+1)$-TQFTs are discrete gauge theories and $BF$ theories.  Both families are related to the Crane-Frenkel-Yetter (CFY) $(3+1)$-TQFTs based on unitary pre-modular categories, which can be realized by lattice models.

A more general framework for the CFY TQFTs are the Mackaay's TQFTs based on spherical $2$-categories \cite {Mac}. Cui generalized the CFY construction to $G$-crossed braided fusion categories \cite{Cui17}.  The resulting new $(3+1)$-TQFTs do not fit into Mackaay's notion of spherical $2$-categories.  Therefore, one problem is to formulate a higher category theory that underlies all these theories, and study their application in $3D$ TPMs.  Cui's TQFTs can also be realized with exactly solvable lattice models \cite{WW17}.

\subsection{Crane-Frenkel-Yetter TQFTs and Tetra-categories}

The lattice models for $(3+1)$-TQFTs are generalizations of the LW models in $2D$.  We expect a tetra-category, which is some double of the input tri-category, to describe all elementary excitations.  Algebraic formulation of general tri- and tetra- categories are extremely complicated.  The physical intuition we gain from the lattice models would help us to understand these special tetra-categories.

\subsection{Representations of Motion Groups and Statistics of Extended Objects}

One lesson that we learned from $2D$ is that deep information about TQFTs is encoded in their associated representations of the mapping class groups, in particular the braid groups.  The braid groups are motion groups of points in the $2D$ disk.  In $3D$, given a $3$-manifold $M$ and a (not necessarily connected) sub-manifold $N$, we can define the motion group of $N$ in $M$.  The first interesting cases will be for links in $S^3$. While the partition functions of the CFY $(3+1)$-TQFTs are not necessarily interesting topological invariants, their induced representations of the motion groups could be more interesting.

The Mueger center describes the pointed excitations in lattice models.
More interesting in 3D are the loop excitations or in general excitations of the shape of any closed surface.  Consider the simplest case for $n$ unlinked unknotted oriented closed loops in $\R^3$, we obtain the $n$-component loop braid group $\LB_n$.  The elementary exchange of two loops leads to a subgroup isomorphic to the permutation group $S_n$, and the passing-through operation to a subgroup isomorphic to the braid group $\B_n$.  It follows that the loop braid group for $n$ unlinked unknotted oriented closed loops in $\R^3$ is some product of the permutation group $S_n$ with the braid group $\B_n$.

By general TQFT properties, each Crane-Frenkel-Yetter TQFT will lead to a representation of the loop braid group.
Physically, we are computing the generalized statistics of loop excitations in the lattice model.  There is no reason to consider only unknotted loops. Similarly we can consider statistics of knotted loop excitations.

\subsection{Fracton physics}

Another question in $4d$ physics is how topological is the fracton physical systems \cite{Haah, VHF}.  The low energy effective theories of fracton systems are not TQFTs as the ground state degeneracy of a fixed space manifold grows as the lattice size grows.  Therefore, a new framework beyond anyons is necessary to capture these new systems.

\section{Applications}

An immediate application of topological phases of matter is the construction of a scalable fault-tolerant quantum computer.  In quantum computation, qubit is an abstraction of all $2$-level quantum systems, though it is not the same as any particular one just like the number $1$ is not the same as an apple.  Quantumness should be a new source of energy, and a  quantum computer is a machine that converts quantum resources such as superposition and entanglement into useful energy.  We need new constants similar to the Planck constant or Boltzmann constant to quantify the energy in superposition and entanglement.

\end{document}